\documentclass[pra,aps,twocolumn,amsmath,amssymb]{revtex4}
\usepackage{graphicx}% Include figure files
\usepackage{dcolumn}% Align table columns on decimal point
\usepackage{bm}% bold math
\usepackage[extension=xxx]{hyperref}
\def\hb{\hbar}
\newcommand{\beq}{\begin{equation}}
\newcommand{\eeq}{\end{equation}}
\newcommand{\beqa}{\begin{eqnarray}}
\newcommand{\eeqa}{\end{eqnarray}}
\def\ra{\rangle}
\def\la{\langle}\def\beq{\begin{equation}}

\usepackage{color}
\begin{document}

\title{Reduction of local velocity spreads by linear potentials}
\date{\today}

\author{F. Damon$^{1,2}$}
\author{F. Vermersch$^{3,4}$}
\author{J. G. Muga$^{3,4,5}$}
\author{D. Gu\'ery-Odelin$^{3,4}$}

\affiliation{$^{1}$ Universit\'e de Toulouse ; UPS ; Laboratoire de Physique Th\'eorique, IRSAMC ; F-31062 Toulouse, France} 
\affiliation{$^{2}$ CNRS ; UMR 5152 ; F-31062 Toulouse, France}
\affiliation{$^{3}$ Universit\'e de Toulouse ; UPS ; Laboratoire Collisions Agr\'egats R\'eactivit\'e, IRSAMC ; F-31062 Toulouse, France} 
\affiliation{$^{4}$ CNRS ; UMR 5589 ; F-31062 Toulouse, France}
\affiliation{$^{5}$ Departamento de Qu'mica F'sica, Universidad del Pa's Vasco, Euskal Herriko Unibertsitatea, Apartado 644, Bilbao, Spain}

\begin{abstract}
We study the spreading of the wave function of a Bose-Einstein condensate accelerated by a constant force both in the absence and in the presence of atom-atom interactions. We show that, despite the initial velocity dispersion, the local velocity dispersion defined at a given position downward can reach ultralow values and be used to probe very narrow energetic structures. We explain how one can define quantum mechanically and without ambiguities the different velocity moments at a given position by extension of their classical counterparts. We provide a common theoretical framework for interacting and non-interacting regimes based on the Wigner transform of the initial wave function that encapsulates the dynamics in a scaling parameter. In the absence of interaction, our approach is exact. Using a numerical simulation of the 1D Gross-Pitaevskii equation, we provide the range of validity of our scaling approach and find a very good agreement in the Thomas-Fermi regime. We apply this 
 approach to the study of the scattering of a matter wave packet on a double barrier potential. We show that a Fabry-Perot resonance in such a cavity with an energy width below the pK range can be probed in this manner. We show that our approach can be readily transposed to a large class of many-body quantum systems that exhibit self-similar dynamics.
\end{abstract}
\maketitle
\section{Introduction}

%GONZALO:REFS Cohen,Real,SPM ARE NOT CITED THEY SHOULD PROBABLY BE INCLUDED COULD YOU PUT THEM AT THE RIGHT PLACE ?

Narrowing down the velocity spread of a packet of particles or a beam is important
in many techniques and experiments: to induce,  control or measure energy-dependent phenomena, 
such as sharp resonances or energy thresholds;  to probe, by scattering, surfaces or interaction potentials with high energy resolution; or to improve   interferometers, spectroscopic studies,  and metrological 
devices such as atomic clocks.  
A recent application example is the study of very low-energy molecular collisions thanks to an accurate velocity control of merged supersonic molecular beams \cite{Ed12}.  

Frequently the particles are produced at a preparation chamber and interact with a material target or field 
elsewhere after a free or guided  flight. This means that the velocity spread of interest is the one of the particles as they arrive at the 
interaction region. The standard (global) velocity spread of the ensemble of particles, irrespective of their location, is only relevant   inasmuch as it affects the local velocity spread, which, 
as we shall see, may differ substantially from the global one. 

In this paper we propose and characterize a method to achieve a narrow local velocity distribution
for one dimensional motion. Our main motivation is
to set high resolution collision experiments between ultracold atoms and a well localized obstacle, but the principles involved are applicable beyond that goal.  
Local velocities and their statistics may be defined in classical mechanics since  trajectories 
carry simultaneous position and velocity 
information, but in  quantum mechanics position and momentum do not commute. 
This does not mean though that a quantum local velocity is a meaningless concept \cite{Muga98}.   
A detailed analysis of specific detectors to perform a local velocity measurement would be necessary in order to determine the exact operators (i.e. the specific quantization among the many possible) 
and/or Positive Operator Valued Measures (POVMs) involved. This operational approach is out of the scope of this paper. Instead we shall adopt a simple phase-space Wigner-function description of the dynamics and the corresponding Weyl rule of quantization to specify the local quantum observables
starting from classical expressions, so that we shall treat formally quantum and classical systems alike in phase-space. 
In any case 
we shall assume conditions in which the differences among different  quantization rules are  negligible. 

A well-known narrowing of the local velocity distribution occurs for free motion. 
If a cloud of classical (noninteracting) particles with an initial phase-space density without position-momentum correlations and zero average position and velocity is let to expand freely, faster particles will advance beyond the slow ones, so that at 
a distant observation point the particles that arrive {\it at a specific instant of time} have well defined velocities. Thus this type of  free-motion induced velocity narrowing only applies to the instantaneous local velocity distribution.  The initial spread for all particles 
irrespective of their location is
actually constant in time, and implies that the dominant, instantaneous (mean) velocity at the observation  point changes with time. In other words, the instantaneous local velocity spread is small, but the spread in time of the instantaneous local mean velocities may be large. 
This limits severely the applicability and usefulness of free-motion narrowing. If we are interested in a specific velocity range, 
many, even most of the particles, i.e., those that do not arrive at the ``right''  time interval to match our requirements,  should be discarded.       
Instead, we propose a mechanism so that the velocities of all particles irrespective of their arrival time is sharply defined.
It only requires a constant force, which for neutral cold atoms may be implemented by using gravity or a magnetic field gradient on magnetically polarized atoms. 

\begin{figure}[h]
\begin{center}
\includegraphics[height=4.8cm,angle=0]{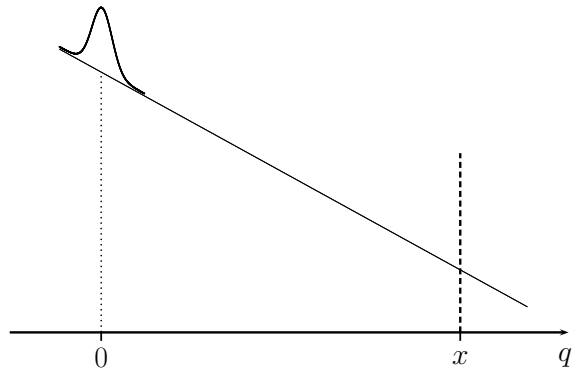}
\end{center}\caption{A wave packet initially located at $q=0$ and subjected to a constant force evolves towards an interacting region located at $q=x$.}
\label{fig1}
\end{figure}

We shall study the experimental situation depicted in Fig.~\ref{fig1}. In  Secs.~\ref{gf} and \ref{gf2}, the general formalism 
to characterize the velocity distribution at the obstacle location $x$ is worked out. 
In Sec. \ref{nig}, we will apply it to noninteracting particles, and in Sec. \ref{TFR} to interacting particles in the Thomas-Fermi regime.   
In Sec.~\ref{secgain}, a full numerical study of the scattering on a double well potential from a carefully outcoupled matterwave is carried out. It shows the possibility to investigate matter wave Fabry-Perot cavities.
\section{General formalism\label{gf}}
The statistical analysis we perform is based on a double average,  over velocity and time, 
at the observation location. To that end 
we follow, interchanging the roles of time and position, the double averaging performed in \cite{Muga98} over position
and momentum at fixed time.   
The definition of the local instantaneous averages over the arriving velocities could in principle
be carried out in several ways with different physical implications.  
For a phase-space distribution $W(q,p; t)$ normalized to one at any time $t$ when integrating over positions $q$ 
and momenta $p$,  a local, instantaneous  average at $q=x$ and time $t$ may be defined as 
\beq
\label{first}
v_x(t)=\frac{\int {\rm d}p\, v\, W(x,p;t)}{\int {\rm d}p W(x,p;t)},
\eeq
where $v=p/m$ and $m$ is the mass of the particle. 
This average is known in hydrodynamical formulations as the ``velocity field'',  see e.g. \cite{PS,book}, 
and it becomes operationally meaningful if the velocities of the particles between $x$ and $x+{\rm d}x$ are probed instantaneously. 
The normalization factor in the denominator is just the particle density divided by the total number of atoms
$N$. 
For many experiments and detectors the particles arrive according to a certain time-dependent flux and 
the relevant quantity is not the number of particles {\it present} in ${\rm d}x$ with a specific velocity, but 
rather the number of particles {\it arriving} (crossing $x$)  with velocities between $v$ and $v+{\rm d}v$ in ${\rm d}t$. 
This is given by $Nv W(x,p,t) {\rm d}v {\rm d}t$ so that the local instantaneous mean velocity (and similarly for higher moments) is defined as   
\beq
\label{first}
\bar{v}_x(t)=\frac{\int {\rm d}p\, v^2 W(x,p;t)}{\int {\rm d}p\, v W(x,p;t)},
\eeq
where the denominator is now the current density $J_x(t)$ or flux (per particle) instead of the density. 
This is in particular the average applicable for scattering experiments \cite{GCG10,dgo11,dgoprl11,dgo12}. 
A basic assumption is that all particles arrive from the left (all velocities are positive at $x$) so  
$J_x(t)\ge0$ and $\int {\rm d}t J_x(t)=1$.

%%%%%%%%%%%%%%%%%%%%%%%%%%%%%%%%%%%%%%%%%%%%%begin figure%%%%%%%%%%%%%%%%%%%%%%%%%%%%%%%%%%%%%%%%%%%%%%%%%%%%%%%%%
%%%%%%
%
Let us first assume a packet of classical non-interacting atoms described by a phase-space-density distribution $W(q,p;t)$ normalized to one. This can be translated into quantum mechanics by means of 
phase-space quasi-distribution functions. 
For the problem at hand (a linear-in-$q$ potential) the Wigner function is a natural choice,
since its propagator for non-interacting particles 
is identical in classical and quantum mechanics \cite{Muga98}. 

The fraction of the number of atoms ${\rm d}^2N^{(v)}_x(t)$ with a velocity between $v$ and $v+{\rm d}v$ that crosses the plane $q=x$ between $t$ and $t+{\rm d}t$ is
\beq
{\rm d}^2N^{(v)}_x(t)=W(x,p;t)v{\rm d}p{\rm d}t\equiv P_x(v,t){\rm d}v{\rm d}t.
\eeq
The probability distribution $P_x(v,t)$ represents the probability density of particles in velocity space per unit time that crosses the plane $x$ at $t$.  We assume that all particles have a positive velocity when crossing $x$, so
\beq
\int\!\!\!\!\!\int\! P_x(v,t){\rm d}v{\rm d}t = \int J_x(t){\rm d}t =1. 
\eeq
The quantity $J_x(t)$ corresponds to the local atomic flux divided by $N$, 
\beq
J_x(t) = \frac{{\rm d}N_x(t)}{{\rm d}t} = \int \frac{{\rm d}^2N^{(v)}_x(t)}{{\rm d}t}  = \int  W(x,p;t)v{\rm d}p, 
\eeq
where $N_x$ is the probability to find a particle at $q>x$. 
In order to define local instantaneous averages it is convenient to introduce the conditional probability density $P_x(v|t)$: 
$P_x(v|t){\rm d}v$ represents the probability  
to find the velocity between $v$ and $v+{\rm d}v$ for the particles that cross $x$ between $t$ and $t+{\rm d}t$. It is normalized to one, 
\beq
\int P_x(v|t) {\rm d}v=1.
\eeq
According to Bayes law, the simple relation
\beq
P_x(v,t) = P_x(v|t) J_x(t)
\eeq
holds. 
The instantaneous mean local velocity at position $x$ and time $t$ is  given by
\beq
\bar v _x(t) = \int v P_x(v|t) {\rm d}v. 
\eeq
Similarly, the instantaneous local mean square velocity reads
\beq
\overline{v^2_x} (t) = \int v^2 P_x(v|t) {\rm d}v. 
\eeq
The mean local velocity is defined as an average over time of the instantaneous mean local velocity,
\beqa
\langle v \rangle_x &=& \int\!\!\!\!\!\int\!  v P_x(v,t) {\rm d}v{\rm d}t = \int\!\!\!\!\!\int\!  v^2 W(x,p;t) {\rm d}p{\rm d}t
\nonumber\\
&=&\int \bar v _x(t) J_x(t) {\rm d}t.
\eeqa
Similarly, the local variance of the velocity reads
\begin{eqnarray}
&&\!\!\!\!\!\!\!\!\!\!\!\!\!\!\!\!(\Delta v_x)^2\!\equiv\!\langle  (v \!-\! \langle v \rangle_x)^2  \rangle_x\!   =\! \! \int\!\!\!\!\!\int\!  (v \!-\! \langle v \rangle_x)^2\! P_x(v,t) {\rm d}v{\rm d}t \nonumber \\
& =&   \int\!\!\!\!\!\int\!  (v - \bar v _x + \bar v _x - \langle v \rangle_x)^2 P_x(v,t) {\rm d}v{\rm d}t 
\nonumber \\
& =&   \int\!\!\!\!\!\int\!\left[ (v - \bar v _x)^2 + (\bar v _x - \langle v \rangle_x)^2 \right]P_x(v,t) {\rm d}v{\rm d}t
\nonumber \\
&=&    \int\!\!\left[ \int\!\! (v\! -\! \bar v_x)^{\!2} P_x(v|t){\rm d}v\! +\! (\bar v _x \!-\! \langle v \rangle_{\!x}\!)^2
\!\right]\!\! J_x(t){\rm d}t\nonumber \\
&=& \int\!\!\left[ \sigma^2_{v|x}(t)+ D^2_{v|x}(t)  \right]\! J_x(t) {\rm d}t,
\label{ecua}
\end{eqnarray}
where
\beqa
\sigma^2_{v|x}(t)& \equiv& \frac{1}{J_x(t)}\int (v - \bar v _x)^2 P_x(v,t){\rm d}v,
\nonumber\\
D^2_{v|x}(t)&\equiv&(\bar v _x - \langle v \rangle_x)^2,
\eeqa
are, respectively, the instantaneous local velocity variance and the squared deviation of the instantaneous local 
mean velocities with respect to the local average velocity.  
According to Eq.~(\ref{ecua}), the variance of the velocity at $x$ is  the sum of two terms: the time average of the instantaneous local variance and the variance of the instantaneous mean velocities. 

Introducing the velocity moments 
\beq
V_n=V_n(x,t) \equiv \int  v^n W(x,{p};t){\rm d}{p},
\eeq
we may rewrite the above quantities as  
\begin{eqnarray}
 J_x(t)& =& V_1(x,t),
\nonumber \\ 
\bar v _x(t) &= &\frac{1}{J_x(t)} \int v^2 W(x,{p},t) {\rm d}p=\frac{V_2}{V_1},
\nonumber \\ 
\overline{v^2_x} (t) &= &\frac{1}{J_x(t)} \int v^3 W(x,{p};t) {\rm d}p=\frac{V_3}{V_1}, 
\nonumber \\ 
\sigma^2_{v|x}(t) &=& \frac{V_3}{V_1}-\left( \frac{V_2}{V_1} \right)^2,
\nonumber \\ 
\langle v \rangle_x&=&  \int\!\!\!\!\!\int\! v^2 W(x,{p},t) {\rm d}p{\rm d}t = \int {\rm  d}t V_2(x,t),
\nonumber\\
D^2_{v|x}(t)&=& \left(\frac{V_2(x,t)}{V_1(x,t)}-\int V_2(x,t') {\rm d}t'\right)^2,
\nonumber
\end{eqnarray}
so computing the first three moments is enough to calculate the local variance of the 
velocity (\ref{ecua}).  The global velocity dispersion irrespective of the particle location is given by
\begin{equation}
\Delta v(t)=\sqrt{\int V_2(x,t){\rm d}x-\left(\int V_1(x,t){\rm d}x\right)^2}. 
\label{globaldv}
\end{equation}
\section{The time-dependent Wigner function and its velocity moments \label{gf2}}
In the following, we study the dynamics of the ground state wave function of a 1D harmonic trap of angular frequency $\omega_0$ after its sudden release on a linear potential. The scaling formalism developed hereafter is applied to a Bose-Einstein condensate (BEC) in the Thomas-Fermi regime \cite{Castin,shly,Imp10} and to the Gaussian wave function for the non-interacting case. However, it is worth noticing that it can be applied to the class of many-body quantum systems that exhibit a self-similar dynamics  \cite{Demler10,ADC11} as justified in Appendix A.
In the presence of a constant acceleration field, $\gamma$, the time-dependent Wigner function, $W(x,p,t)$, can be obtained from the initial Wigner function, $W_0$, through a proper scaling substitution (see Appendix A), 
\begin{equation}
W(x,p,t) = W_0(X,P),
\label{scalingsolut}
\end{equation}
where we have used the following linear transformation
\begin{equation}
\begin{pmatrix}
X\\
P
\end{pmatrix}=\begin{pmatrix}
 1/\alpha & 0\\
-m\dot\alpha &\alpha
\end{pmatrix}\begin{pmatrix}
x-\eta\\
p-m\dot\eta
\end{pmatrix},\label{eqlintrans}
\end{equation}
{with} $\alpha$ and $\eta$ {two} time-dependent functions. The matrix of this transformation generates both a squeezing in momentum space correlated to a shearing in position space. $\eta$ accounts for the center of the packet motion and obeys the Newton equation $\ddot \eta=\gamma$. The scaling factor $\alpha$ obeys a second order differential equation that depends on the system (see Appendix A):
\begin{equation}
\ddot \alpha = \frac{\omega_0^2}{\alpha^p},
\label{eqalpha}
\end{equation}
with $p=3$ for a Gaussian wave function without interactions or a Tonks-Girardeau gas \cite{OHS02,PSO03,MiG05}, and $p=2$ for the wave function of a Bose-Einstein condensate in the Thomas-Fermi regime. The $X$ and $P$ variables are useful to derive the velocity moments in terms of their initial value $V_n^{(0)}(x)$:
\begin{equation}
V_n(x,t)=\int \left[ \frac{P}{m\alpha}+u({X},t)\right]^nW_0\left({X},P \right)\frac{{\rm d}P}{\alpha},
\label{velmom}
\end{equation}
with $u({X},t)=\dot\eta + \dot\alpha {X}$. We find
\begin{eqnarray}
V_0(x,t) & = & 
%+
V_0^{(0)}(X)/\alpha ,
\label{ev0}
\\
V_1(x,t) & = & 
%+
uV_0^{(0)}(X)/\alpha,
\label{ev1}
\\
V_2(x,t) & = & V_2^{(0)}(X)/\alpha^3
+u^2V_0^{(0)}(X)/\alpha,
\label{ev2}
\\
V_3(x,t) & = & 
3uV_2^{(0)}(X)/\alpha^3 +u^3V_0^{(0)}(X)/\alpha.
\label{ev3}
\end{eqnarray}
To obtain this result, we have explicitly used the symmetry property of the initial Wigner function: 
$W_0(x,p)=W_0(x,-p)$. The first three moments are therefore directly deduced from two moments of the initial distribution.
Interestingly, the first two global velocity moments are readily obtained in a form that depends explicitly on the initial condition and the scaling parameters. We find $\langle v \rangle (t) =\dot\eta$ and the velocity dispersion
\begin{equation}
\Delta v^2 (t)  = \frac{\Delta v^2(0)}{\alpha^2} + \frac{2\omega_0^2}{p-1}\left(1-\alpha^{1-p}\right)\Delta x^2(0),\label{deltav}
\end{equation}
using Eq.~(\ref{eqalpha}). With $p=3$, we recover the fact that the velocity dispersion remains constant and equal to its initial value in the case of an interaction-free Gaussian wave packet. With $p=2$ we obtain the explicit expression for the velocity dispersion at any time for a Bose-Einstein condensate in the Thomas-Fermi regime. In particular, its limit for long time is equal to $\Delta v(t \rightarrow \infty) = \sqrt{2}\omega_0 \Delta x(0)$ in agreement with the Virial theorem \cite{RMPSandro,dgopra11}. 
\section{Application to a noninteracting gas\label{nig}}
Let us set the initial phase-space distribution as an uncorrelated product of Gaussians,  
\beq
W_0(x,p)=\frac{1}{2\pi}\frac{1}{\Delta x_0\Delta p_0}{\rm e}^{-x^2/2\Delta x_0^2}{\rm e}^{-p^2/2\Delta p_0^2}, 
\eeq
obeying $\Delta x_0 \Delta p_0 =\hbar/{2}$. 
For a classical ensemble $\hb$ could be an arbitrary constant whereas 
in quantum mechanics this is Planck's constant$/2\pi$.
Using $\sigma\equiv\Delta x_0$  we find
\begin{eqnarray}
V_0^{(0)}(x)  & = & \frac{1}{\sqrt{2\pi}\sigma}{\rm e}^{-x^2/2\sigma^2}, 
\nonumber\\
V_2^{(0)}(x)  & = & \left( \frac{\hbar}{2m\sigma}\right)^2V_0^{(0)}(x) , 
\nonumber\\
\alpha(t)   & = & \sqrt{1+\omega_0^2t^2},  
\nonumber\\
u(x,t)  & = & \gamma t + (x-\gamma t^2/2)\omega_0^2 t /(1+\omega_0^2t^2),
\label{uxt}
\end{eqnarray}
with $\omega_0=\hbar / (2m\sigma^2)$.

The quantities $\bar v _x(t)$, $\overline{v^2_x} (t)$ and $\sigma^2_{v|x}(t)$ involve ratios of the form $V_{n>1}(x,t)/V_1(x,t)$. Using Eqs.~(\ref{ev1}), (\ref{ev2}) and (\ref{ev3}), one realizes that these ratios depend only on $V^{(0)}_{2}(X)/V^{(0)}_0(X)$,
which is a constant for a Gaussian Wigner function. In this way, we find
\begin{eqnarray}
\displaystyle\bar v_x(t)&=\frac{\displaystyle u^2+(u^2t^2+\sigma^2)\omega_0^2}{\displaystyle u(1+\omega_0^2t^2)},\\
\displaystyle\overline{v_x^2}(t)&={\displaystyle u\bar v_x(t)+\frac{2\sigma^2\omega_0^2}{\displaystyle 1+\omega_0^2t^2}}.%\\
%\displaystyle\sigma_{v|x}^2(t)&=\frac{\displaystyle \sigma^2\omega_0^2[u^2+(u^2t^2-\sigma^2)\omega_0^2]}{\displaystyle u^2(1+\omega_0^2t^2)^2}.
\end{eqnarray}
The first two global velocity moments are 
\beqa
\label{firstvm}
&&\langle v \rangle(t)=\!\int\! V_1(x,t){\rm d}x\!=\!\gamma t,
\\
\label{secondvm}
&& \langle v^2 \rangle(t)=\!\int\!\! V_2(x,t){\rm d}x\!=\!\frac{\hb^2}{4 m^2 \sigma^2}\! +\! \gamma^2 t^2\!.
\eeqa
To have a negligible fraction of negative velocities we simply assume 
an observation point at $x{\gg}\sigma$ and $t>0$  
hereafter, in particular in all integrals. 
We now calculate the first velocity moments. They can be written in the form 
\beq
V_j(x, t):= 
  {\rm e}^{-\Phi} f_j,
\label{v1}
\eeq
where the $f_j$ are given in Appendix B and 
\beq
\Phi=\frac{m^2 \sigma^2 (\gamma t^2 - 2 x)^2}{8m^2 \sigma^4 + 2\hb^2 t^2}=
-\frac{x^2(1-\gamma t^2/2x)^2}{2\sigma^2(1+\hbar^2t^2/4m^2\sigma^4)}
\eeq
is the phase that governs the behavior of the moments. 
In particular $\Phi=\Phi'=0$ (the primes denote time derivatives) at $t_c=\sqrt{2x/\gamma}$, which is the time that   
a classical particle takes to reach $x$ if it is initially at rest at the origin. 
The second derivative with respect to time at that instant is 
\beq
\Phi''(t_c)=\frac{16 g m^2 \sigma^2 x}{8m^2 \sigma^4 + 4 \hb^2 x/\gamma}
\eeq
which, keeping $x/\gamma$ constant, i.e., for a fixed $t_c$, grows with $x$. In other words, the 
moments become narrower functions of $t$.

We may now calculate the local  
variance as
\beqa
(\Delta v_x)^2=\int_0^\infty {\rm d}t V_3- \left[\int_0^\infty {\rm d}t V_2\right]^2.
\eeqa
Laplace's (or steepest descent) approximation  gives for the integrals 
$\int {\rm e}^{-\Phi} f {\rm d}t \sim 
{\rm e}^{-\Phi(t_c)} f(t_c)\sqrt{2 \pi/\Phi''(t_c)}$ and 
$(\Delta v_x)^2\sim 0$ in sharp contrast with the fixed-time (and time independent) 
global velocity variance $\langle v^2 \rangle(t) - (\langle v \rangle (t) )^2=\hb^2/(4 m^2 \sigma^2)$
calculated with the Wigner function, see Eqs. (\ref{firstvm}) and (\ref{secondvm}). 
%%%%%%%%%%%%%%%%%%%%%%%%%%%%%%%%%%%%%%%%%%%%%begin figure%%%%%%%%%%%%%%%%%%%%%%%%%%%%%%%%%%%%%%%%%%%%%%%%%%%%%%%%%
\begin{figure}[h]
\begin{center}
\includegraphics[height=4.cm,angle=0]{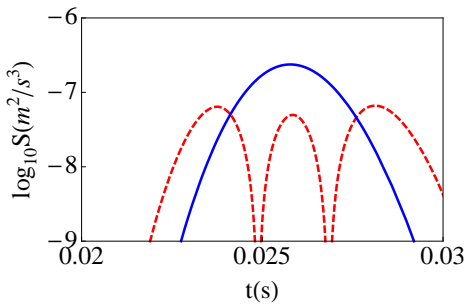}
\caption{\label{vc2}
(Color online) Components of the integrand for the local variance $\Delta v_x^2$:   
$S=\sigma_{v|x}^2(t)J_x(t)$  (dashed red line),  
and $S=D_{v|x}^2(t)J_x(t$) (solid blue line). Parameters: $\gamma=0.6$ m/s$^2$, $x=201$ $\mu$m, mass of rubidium-87.
The initial state is the ground state of a harmonic trap with angular frequency $\omega_0 = 2\pi \times 150$ Hz. Ê  
The passage of a classical particle released from the origin at $t=0$ is at $t_c=0.026$ s.
}
\end{center}
\end{figure}
%%%%%%
In Fig.  \ref{vc2} we show the two components that, when integrated,  make $(\Delta v_x)^2$, see Eq.~(\ref{ecua}): In our configuration 
the local velocity variance  is  dominated 
by the squared deviation of the local instantaneous mean.   
The parameters are chosen as specified in the caption
for realistic conditions on initial velocity width, acceleration, and observation distance.    
%GONZALO COULD YOU ADAPT THE NUMERICAL VALUES WITH THOSE USED IN THE FOLLOWING FOR SAKE OF CONSISTENCY INCLUDING IN THE FIGURE 
For these parameters  the local variance
$(\Delta v_x)^2(0.5$mm)$=8.22\times10^{-10}$ (m/s)$^2$, is already more than two orders of magnitude smaller than 
the global variance $(\Delta v)^2=3.4\times 10^{-7}$ (m/s)$^2$. 

The two variances (global and local) are quite different because of their 
distinct physical content. $(\Delta v)^2$ is independent of $x$, $\gamma$  or $t$
and corresponds to measuring the velocities of all the particles of the statistical ensemble, wherever they are, 
at a given instant. Since all particles are equally accelerated their 
initial velocity differences remain the same at any time. $(\Delta v)_x^2$ results instead from local velocity
measurements weighted by $P_x(v,t)$. This distribution can be narrowly peaked when  
the velocities at $x$ are predominantly due to the common effect of $gx$ rather than to the effect of the initial spread of the initial conditions. 
Specifically we have 
that, at position $x$, the velocity of a particle that at time $0$ was at $x_0$ with velocity $v_0$, 
is 
\beq
v_x=\sqrt{\gamma(x-x_0)+v_0^2},
\eeq
i.e., any difference due to the the initial conditions is washed out for a large $x\gamma$, 
which leads to  velocity uniformity.  The conditions 
$x{\gg}\sigma$ and $x \gamma{\gg}\sigma_v^2$, where $\sigma_v^2$ is the initial variance of the velocity, 
guarantee the dominance of the first term and lead to a local velocity spread suppression.
Note that, when $\sigma$ and $\sigma_v$ are related by the minimum uncertainty product relation,  i.e., 
$\sigma_v=\hbar/(2m\sigma)$, very small or very big $\sigma$ values 
require a distant $x$ 
to wash out the differences in the arriving velocities.   
Indeed, all previous results can be translated  formally into quantum mechanics
by interpreting $W$ as the Wigner function and the dynamical variables 
according to Weyl's quantization rule.  
In  the Weyl-Wigner framework  
the dynamical variable $\delta(x-q)v$ corresponds to the  
flux operator 
$\hat{J}_x=(1/2)(\hat{v}\delta(x-\hat{q})+\delta(x-\hat{q})\hat{v})$ 
and, more generally, $v^n\delta(x-q)$ to hermitian operators
with matrix elements 
\beq
\la p|O^{(n)}|p'\ra=\left(\frac{p+p'}{2m}\right)^n\la p|x\ra\la x|p'\ra 
\eeq
in momentum space. 
As other quantization rules are possible, the above results, which are based on one of them, 
may be questioned because of their non-uniqueness. 
However, 
actual measurements typically 
imply an averaging 
that tends to smooth the differences among the different rules.  
Moreover the conditions of the experiment discussed are such that at each instant the local quantum field is
essentially monochromatic. This again washes out differences among quantization rules:  as interferences 
do not play any major role for our setting and the effects of the non-commutativity of position and momentum are negligible, 
a classical-like  interpretation of the results is justified. This is in fact the usual approach  for most  
time-of-flight experiments.     
\section{Application to a Bose-Einstein condensate in the Thomas-Fermi regime\label{TFR}}  
Unlike the Gaussian case, the Wigner function of the wave function associated with a Bose-Einstein condensate in the Thomas-Fermi regime cannot be worked out analytically. In this section, we compare numerics with the scaling approach to obtain the different velocity moments accurately. The parabolic profile of the condensate in the Thomas-Fermi regime has been used to deduce Eqs.~(\ref{scalingsolut}) and (\ref{eqalpha})(see Appendix A) that gives the time evolution of the scaling parameter. However, a numerical study is necessary to validate the determination of the velocity moments with the scaling approach since the parabolic profile does not encapsulate the smooth edge of the wavefunction \cite{dalfovo}. We show in the following that the knowledge of the initial Wigner function computed numerically is sufficient to obtain all required information on the dynamics and in particular on the different velocity moments with a high accuracy.

The initial wave function is found by evolving the 1D time-dependent Gross-Pitaevskii (GP) equation in imaginary time using a split-step Fourier method \cite{MA09}. In practice, we use an initial Gaussian trial wavefunction with a mean quadratic size evaluated from the expected parabola shape and we iterate until the final wavefunction is stationary with respect to real-time evolution. The mean-field wave function corresponds to that of a Bose-Einstein condensate with $N$ rubidium-87 atoms (scattering length $a_{\rm sc}=5$ nm) held in a harmonic trap of angular frequency $\omega_0=2\pi\times 150$ Hz. 
The Thomas-Fermi regime is reached when $\chi=N a_{\rm sc}/a_0\gg1$ where $a_0=(\hbar/m\omega_0)^{1/2}$ is the length scale associated with the harmonic confinement ($\chi=1$ corresponds to $N\approx176$ atoms).

\begin{figure}[h!]
\begin{center}
\includegraphics[width=0.95\linewidth,angle=0]{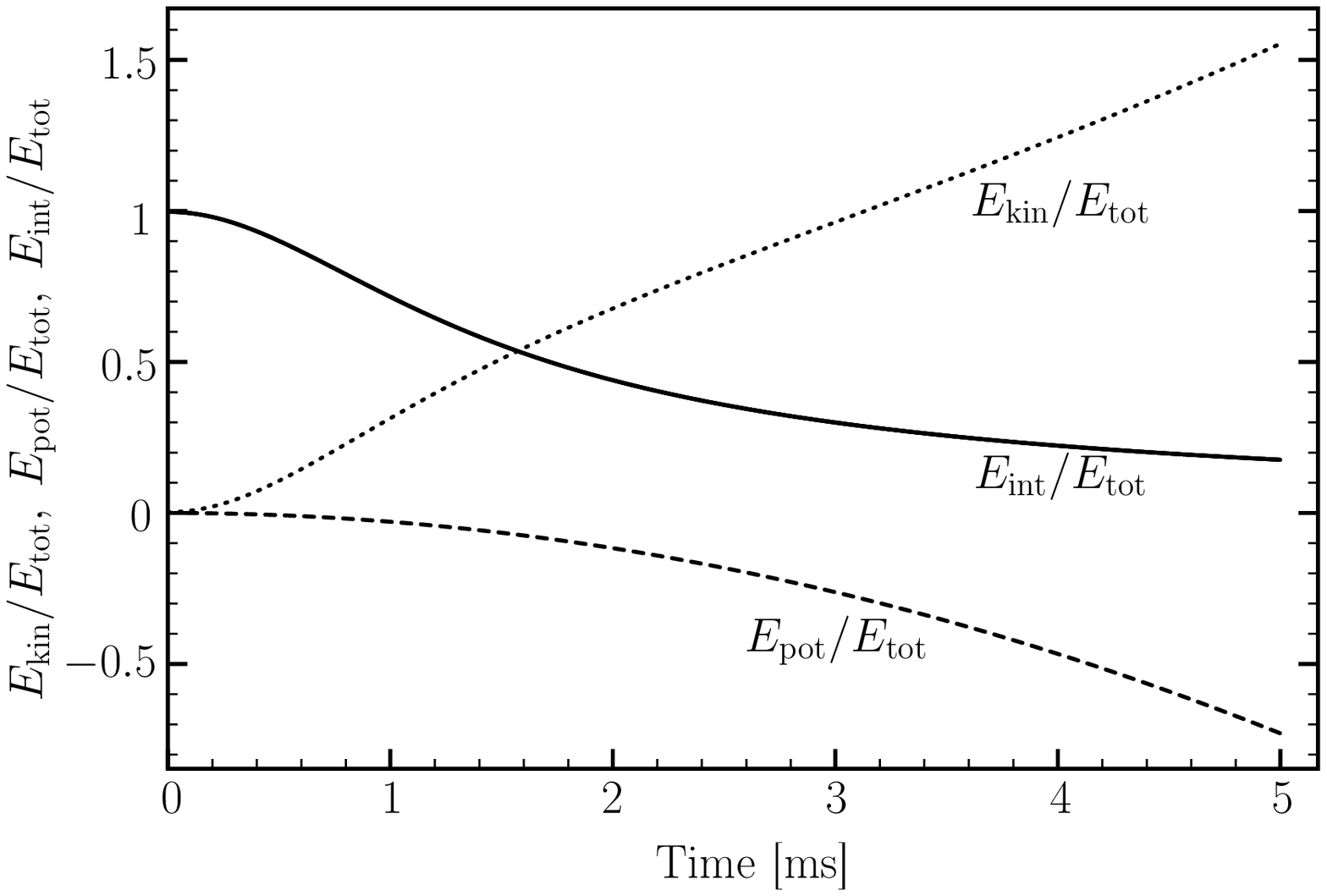}
\end{center}
\caption{Evolution of the kinetic ($E_{\rm kin}$, dotted line), potential ($E_{\rm pot}$, dashed line) and interaction energy ($E_{\rm int}$, solid line) of a Bose-Einstein condensate as a function of time during its propagation over the slope (energies are normalized to the total energy $E_{\rm tot}=E_{\rm kin}+E_{\rm pot}+E_{\rm int}$) . Parameters: the initial wave function corresponds to that of a Bose-Einstein condensate of rubidium-87 in the Thomas-Fermi regime with $N=17600$ atoms initially at equilibrium in an harmonic potential of angular frequency $\omega_0 = 2\pi \times 150$ Hz. At time $t=0$, the trapping potential is suddenly removed and the packet experiences the acceleration ($\gamma=0.6$ m/s$^2$) due to the linear potential.}
\label{fig3}
\end{figure}

\begin{figure}[h!]
\begin{center}
\includegraphics[width=0.95\linewidth,angle=0]{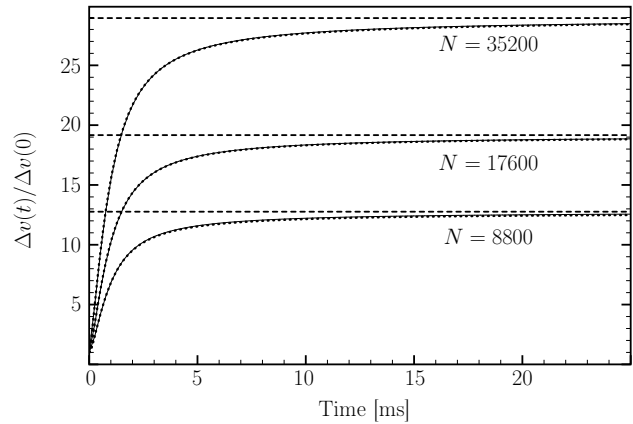}
\end{center}
\caption{Time evolution of the global velocity dispersion $\Delta v(t)/\Delta v(0)$ of a Bose-Einstein condensate for three different values of the atom number $N=8800,\,17600,\, 35200$. Solid lines: theoretical prediction using the scaling approach (see Eq.~\ref{deltav}), dotted lines numerical integration of the Gross-Pitaevskii equation, dashed line: theoretical asymptote at long time.}
\label{fig4}
\end{figure}

In the following, we consider the time evolution of the wave function resulting from an abrupt release of the condensate on the slope ($V(x)=m\gamma x$, $\gamma=0.6$ m/s$^2$), the case of a progressive outcoupling is addressed in Sec.~\ref{secgain}. We compare the results obtained by a full integration of the Gross-Pitaevskii equation with those resulting from the scaling approach (Eqs.~(\ref{deltav}) and (\ref{eqalpha})). In the course of the propagation, the interaction energy is converted into kinetic energy as illustrated in Fig.~\ref{fig3}, and more precisely it is responsible for the increase of velocity dispersion. After $5$ ms of propagation for a BEC initially in the Thomas-Fermi regime, already 80 $\%$ of the interaction energy has been transferred. Figure \ref{fig4} shows the evolution of the velocity dispersion depending on the number of atoms. The impressive agreement demonstrates the efficiency of the scaling approach.

\begin{figure}[h!]
\begin{center}
\includegraphics[width=0.95\linewidth,angle=0]{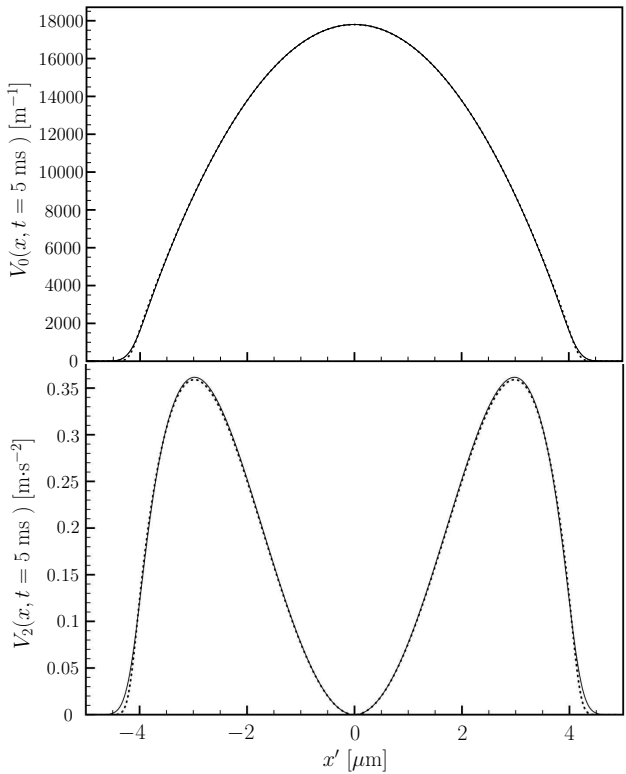}
\label{compar}
\end{center}
\caption{Velocity moments $V_0(x,t=5\text{ ms})$ (density profile) and $V_2(x,t=5\text{ ms})$ for $N=35200$ atoms, where the positions are expressed in the center-of-mass coordinates $x'=x-\eta$: full numerical procedure including the time evolution (solid line), result of the scaling approach based only on the initial Wigner function (dotted line).}
\label{fig5}
\end{figure}

The numerical integration of the GP equation with split-Fourier method is quite efficient even though the time step should be smaller for large-size wavefunction (large $N$). However, the computation of the Wigner function turns out to be rapidly cumbersome since the required grid in phase-space has a size that should increase with time to guarantee a good accuracy. We now compare the first two non-zero velocity moments $V_0(x,t)$ and $V_2(x,t)$ obtained after 5 ms of time evolution on the slope from the full numerical procedure of the dynamics (integration of the Gross-Pitaevskii equation $+$ calculation of the Wigner function at time $t=5$ ms) with the result of the scaling approach for which we use only the computing of the initial Wigner function combined with the time evolution of the scaling parameter (Eqs.~ (\ref{eqlintrans}) and (\ref{eqalpha})). Results are summarized in Fig.~\ref{fig5} for $N=35200$ atoms. %The larger $N$ the better the agreement including in the wings of the 
 %wave function.

To compare quantitatively the velocity moments, we define the relative error after 5 ms of propagation: 
\begin{equation}
\varepsilon(N)=\left|1-\frac{\Delta_2 v(t=5\text{ ms})}{\Delta_1 v(t=5\text{ ms})} \right|
\end{equation}
in which $\Delta_1 v$ is the velocity dispersion extracted by the full numerical approach, and $\Delta_2 v$ is the one resulting from the rescaled initial Wigner function. Results are summarized in the following table:
\begin{center}
\begin{tabular}{|c | c c c c c c|}
\hline
$N$ & 176 & 880 & 1760 & 8800 & 17600 & 35200\\
\hline\hline
$\varepsilon$ & 0.278  & 0.0741	& 0.0397 & 0.0132	& 0.0104	& 0.009	 \\ \hline
\end{tabular}

{\small Table 1: Numerical values of $\varepsilon(N)$.}\label{table1}
\end{center}

\begin{figure}[h!]
\begin{center}
\includegraphics[width=0.95\linewidth,angle=0]{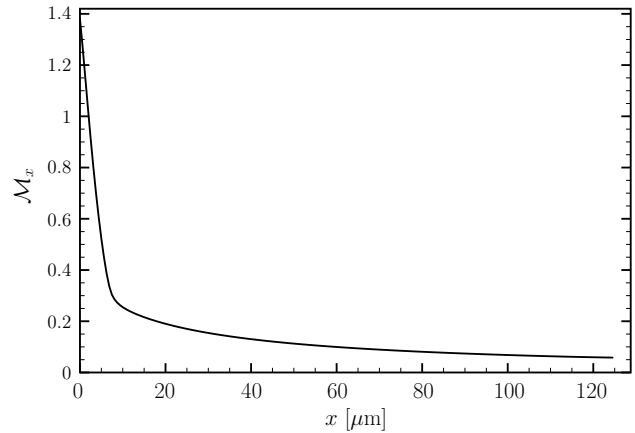}
\end{center}
\caption{Local monochromaticity as a function of the distance $x$ at which it is measured. This data has been obtained for a Bose-Einstein condensate in the Thomas-Fermi regime with $N=35200$ atoms, other parameters are identical to those of Fig.~\ref{fig3}.}
\label{monochrom}
\end{figure}

The larger the interaction parameter $\chi$ ($\propto N$), the better the agreement. We conclude that our scaling approach gives a good account of the evolution of the wave function of a Bose-Einstein condensate in the Thomas-Fermi regime. As the validity of our approach is now well demonstrated, we can extend its use for much larger time. We introduce in the following the quantity $\mathcal{M}_x$ which characterized the improvement of the local monochromaticity as a function of the distance from the original trap:
\begin{equation}
\mathcal{M}_x=\frac{\Delta v_x}{\langle v\rangle_x}=\frac{\sqrt{\int_0^\infty V_3(x,t){\rm d}t-\left(\int_0^\infty V_2(x,t){\rm d}t\right)^2}}{\int_0^\infty V_2(x,t){\rm d}t}.
\end{equation}
To calculate the moments involved in the local monochromaticity quantity, $\mathcal{M}_x$, we use the scaling laws provided by Eqs.~(\ref{ev2}) and (\ref{ev3}). The evolution of this quantity as a function of the distance is plotted in Fig.~\ref{monochrom}. We recover here quantitatively the main result of this section, the local monochromaticity increases with the distance even for a many-body wave function with large repulsive interaction. This is to be contrasted with the global velocity dispersion which increases as shown in Fig.~\ref{fig4}. An interacting BEC evolving in the presence of a constant force can therefore be used as a local probe in a scattering experiment. In the following section, we give a concrete example in which the outcoupling of the wave function from its original trap is performed progressively to amplify the local monochromaticity including in the case of a Bose-Einstein condensate in the Thomas-Fermi regime.

\section{Gain obtained from a progressive outcoupling}  
\label{secgain}

\begin{figure}[h]
\begin{center}
\includegraphics[height=8.8cm,angle=0]{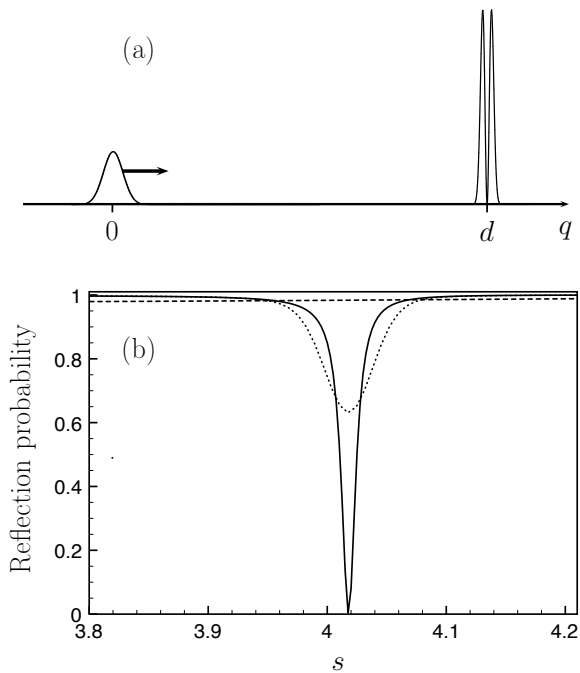}
\caption{\label{plat}
Matter wave probing of a Fabry-Perot resonance. The reflection probability is plotted as a function of the double barrier height: for an incoming plane wave (solid line) with the velocity $v_0$ that coincides with the resonance ($s=4.02$), for an incident wave packet with a velocity dispersion of 600 $\mu$m/s (dashed line) and 20 $\mu$m/s (dotted line).}
\end{center}
\end{figure}

\begin{figure}[h]
\begin{center}
\includegraphics[height=8.8cm,angle=0]{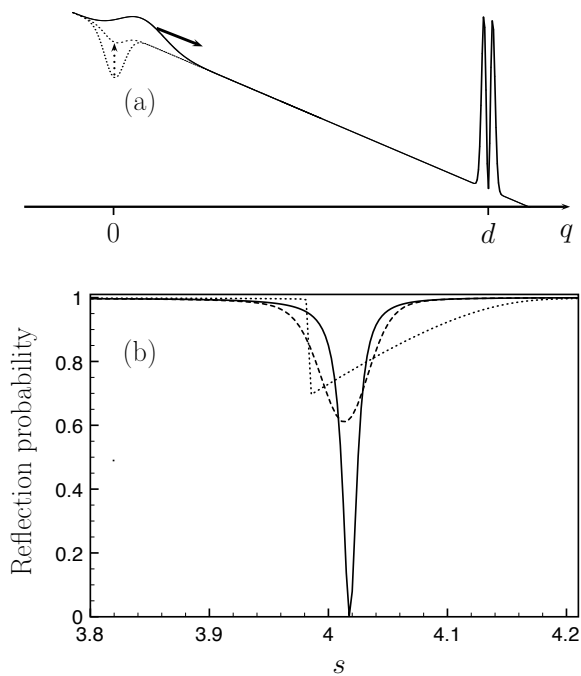}
\caption{\label{pente}
Matter wave probing of a Fabry-Perot resonance placed on a slope. The reflection probability is plotted as a function of the double barrier height: for an incoming plane wave (solid line) with the velocity $v_0$ that coincides with the resonance ($s=4.02$), for an incident wave packet with a velocity dispersion of 600 $\mu$m/s outcoupled in 80 ms (dashed line) and for a Bose-Einstein condensate in the Thomas-Fermi of velocity dispersion 2 mm/s outcoupled in 100 ms (dotted line).
}
\end{center}
\end{figure}

In this section, we compare the matter wave probing of a very thin Fabry-Perot resonance associated with a repulsive double barrier \cite{Wilkens93,Billy07} using two different protocols: first we consider a wave packet at a constant velocity with a well-defined velocity dispersion and in the absence of other extra external potential (see Fig.~\ref{plat}.a); second we consider a situation in which the matter wave is accelerated by a linear potential towards the double barrier (see Fig.~\ref{pente}.b) and for which atoms can also be progressively outcoupled from their original trap \cite{Bloch99,weitz03,RSH04,RMH05,RFH06}.

The double barrier potential is modeled by a sinusoidal potential located at $x=d$:
\begin{eqnarray}
U_{\rm db}(x) & = & 2sE_R\left( 1 - \cos(2\pi (x-d) / d_R) \right)  \nonumber\\
& \times & \Theta(x-d+d_R)\Theta(d+d_R-x),
\label{potd}
\end{eqnarray}
where $E_R=h^2/(2md^2_R)=mv_R^2/2$, $\Theta$ is the Heaviside step function ($\Theta(x<0)=0$ and $\Theta(x\ge 0)=1$). The numerical simulations presented hereafter are performed with the realistic values $d=201$ $\mu$m, $d_R=0.48$ $\mu$m, $v_R=h/(md_R)=9.565$ mm/s \cite{dgopra09,dgoprl11}. The Fabry-Perot resonance that is probed has an energy $E_0=mv_0^2/2$ with $v_0= 1.62 v_R= 15.53 $ mm/s for a barrier height of $16.08E_R=6.13E_0$. The width of this resonance is $E_0/1000$ (i.e. 0.36 pK in temperature units!) that corresponds to a velocity dispersion 8.3 $\mu$m/s. For comparison, let us recall that the typical order of magnitude of the velocity dispersion achievable with a rubidium BEC is $~$2 mm/s \cite{dgoprl11}. 

To scan the resonance we vary the height of the double barrier by adjusting the parameter $s$ for a fixed incoming velocity. The result obtained by considering an incident plane wave of wave vector $k_0=mv_0/\hbar$ is plotted as a reference by a solid line in Figs.~\ref{plat}(b) and \ref{pente}(b). The resonance appears at  $s=4.02$. Interestingly, the width in energy of the reflection probability as a function of $s$ is significantly larger than that of the resonance. Indeed, from a Gaussian fit of the reflection probability, we find a variance in $s$ equal to $\Delta s=0.008$. This means that experimentally if the double barrier is realized by optical means, the intensity should be stabilized at a level better than one per thousand, and not much better than the resonance width $E_0/1000$. This feature facilitates the detection of the resonance.

Figure \ref{plat}(b) contains also the reflection probability about the resonance for an incident wave packet of mean velocity $v_0$ and with a velocity dispersion $\Delta v=$ 600 $\mu$m/s (dashed line) and 20 $\mu$m/s (dotted line). The probability reflection of an incident packet with a velocity dispersion already more than three times smaller than that of a BEC remains very close to one. This is to be contrasted with the case of a packet that have a velocity dispersion 100 times smaller than that of a BEC (i.e. on the order of the width of the resonance). In this latter case, the presence of the resonance can be clearly seen as expected since about 38 \% of the atoms are reflected.

%%%%%%%%%%%%%%%%%%%%%%%%%%%%%%%%%%%%%%%%%%%%%begin figure%%%%%%%%%%%%%%%%%%%%%%%%%%%%%%%%%%%%%%%%%%%%%%%%%%%%%%%%%

We now consider the probing of the same resonance by a wave packet that is accelerated by a constant force from its original trap located at $x=0$. The potential energy experienced by the atoms now reads (see Fig.~\ref{pente}(a))
\begin{equation}
U(x,t)=m\gamma (x-d) - U_0e^{-t/\tau}e^{-2x^2/w_0^2} + U_{\rm db}(x).
\end{equation}
The Gaussian term accounts for the  trap potential which accommodates the initial wave function. The exponential form for the time dependence of the trap depth was used in Ref.~\cite{dgoepl08} to outcouple progressively atoms from a Bose-Einstein condensate into a guide.
The slope has been chosen so that a particle launched with a zero velocity at $x=0$ arrives at $x=d$ with a velocity $v_0$ i.e.~$\gamma=v_0^2/(2d)= 0.6$ m/s$^2$ i.e.~an energy that coincides with that of the Fabry-Perot resonance. 

The analytical formalism developed in the previous sections is the appropriate one for $\tau \rightarrow 0$, i.e. for an instantaneous release of the wave function from its originating trap and a subsequent propagation on the slope. We have already shown that the position-velocity correlation that builds up during such an evolution generates a small local velocity dispersion downward. As demonstrated below on a specific example, an even better strategy consists in releasing the atoms from the trap progressively as done for the generation of guided atom lasers \cite{aaprl06,dgoepl08,dgopra09,arlt12,robins2013}. In such a protocol, the initial conditions of each atom is nearly the same i.e. atoms are released nearly at the same position with a velocity very close to zero. This favors a very thin local velocity dispersion after propagation.

%%%%%%
%%%%%%

We illustrate this idea with the dashed curve of Fig.~\ref{pente}.b that corresponds to the reflection probability of a wave packet (without atom-atom interactions) with an initial velocity dispersion equal to 600 $\mu$m/s and that has been outcoupled in $\tau=80$ ms. About 40 \% of the atoms are reflected when the height of the double barrier is varied. In contrast with the scattering on the double barrier, in the absence of slope, of a packet having the same velocity dispersion, the thin Fabry resonance can here be probed with accuracy. This illustrates clearly the importance of both a progressive outcoupling and the building up of position-velocity correlation in the course of the propagation on the slope. 

We have also performed the progressive outcoupling (over 100 ms) of an interacting BEC with a velocity dispersion even larger on the order of 2 mm/s \cite{dgopra11}. Note that the long outcoupling times are also optimal to reduce the contamination of the transverse excited states in real systems \cite{dgoepl08, dgopra09,dgopra11,arlt12}. The result is plotted as a dotted line in Fig.~\ref{pente}(b). The reflection probability is about 30 \%. Once again the initial source is very far from an ideal one and has a ``very'' large instantaneous velocity width. The situation may appear as even worse since the interaction energy is converted into kinetic energy in the course of the propagation as already discussed in Sec.~\ref{TFR}. However, the benefit of the building up of the position-velocity correlation is dominant. An instantaneous outcoupling of the BEC yields a 10 \% reflection probability, a value that would correspond in the absence of a slope to that of a wave packet having a dispersion velocity on the order of 50 $\mu$m/s, that is 40 times smaller than that of the BEC. An extra gain of a factor three on the reflection probability is here obtained from the progressive outcoupling. Interestingly, this technique avoids completely any limitation in energy dictated by the chemical potential. The shape of the reflection probability as a function of the height of the double barrier is radically different in the case of an interacting BEC compared to that obtained in the absence of atom-atom interactions (compare dotted and dashed lines in Fig.~\ref{pente}.b). This results from the sharp border of the inverse parabolic shape of the BEC assumed in the Thomas-Fermi regime and its conversion in position-velocity correlation in the course of the propagation.

In conclusion, we have shown how it is possible to decrease the velocity dispersion at a given position using a linear potential. This method works efficiently also in the presence of repulsive interaction. It provides a method to circumvent the limitations that may appear in outcoupling processes \cite{Billy07,BillyNLP} and to improve the local monochromaticity of a matter wave source.

In Ref.~\cite{iacopo}, the author proposed to investigate an atom-blockade effect with a Fabry-Perot cavity in close analogy with the Coulomb blockade effect of electronic transport.
An incident atom laser with a very thin velocity dispersion would be reflected or not from a Fabry-Perot cavity depending on the presence or not of an atom in the cavity. Such an experiment appeared as out of reach for a long time because of the very small velocity dispersion ($< 50$ $\mu$m/s) required for its demonstration. Exploiting the position-velocity correlation that builds up in the propagation over a slope we have shown that this level of accuracy can now be achieved with the current state of the art. This work therefore paves the way for ultrahigh resolution in matter wave probing, the observation of nonlinear atom optical effects and of the interplay between atom tunneling and atom-atom interactions \cite{arimondo07,bloch11}. 

We are grateful to B. Georgeot, A. del Campo and J. Billy for useful comments.
This work was supported by Programme Investissements d'Avenir under the program ANR-11-IDEX-0002-02, reference ANR-10-LABX-0037-NEXT, the Institut Universitaire de France,   the Basque Government (Grant IT472-10), MINECO   
(Grant FIS2012-36673-C03-01), and the program UFI 11/55 of UPV/EHU.  
\appendix
\section{Scaling Wigner solution}

This appendix justifies the scaling form of the Wigner function of Eq.~(\ref{scalingsolut}) for both a one-body wave function and a Bose-Einstein condensate in the Thomas-Fermi regime.

We consider a cloud of atoms in a harmonic trap. At time $t=0$, the confinement is suddenly removed and the packet experiences a uniform accelerating force. 
For  a non-interacting gas,  the initial Wigner function, $W_0(x,p)$, obeys the stationary Liouville  equation
\begin{equation}
p\partial_x W_0 - m^2\omega_0^2x\partial_pW_0=0, 
\label{eqq1}
\end{equation}
Multiplying this by $p/N$, where $N$ is the total number of particles, and integrating over $p$ yields
\begin{equation}
\partial_x\left( \int p^2 W_0 {\rm d}p \right) + m^2\omega_0^2 xn_0(x) =0,
\label{eqq2}
\end{equation}
where $n_0(x)$ is the linear density defined by
\begin{equation}
 n_0(x)=N\int  W_0(x,p) {\rm d}p\;\;{\rm with}\;\;\int n_0(x) {\rm d}x=N.
\end{equation}
Multiplying Eq.~(\ref{eqq2}) by $x/N$ and integrating over $x$ provides the usual
momentum relation associated to the harmonic oscillator $\langle p^2\rangle_0 = m^2\omega_0^2 \langle x^2 \rangle_0$. 

For $t>0$, the packet is released on a slope i.e. on a linear potential $U(x)=-m\gamma x$. The equation of evolution of the time-dependent Wigner function now reads 
\begin{equation}
\partial_tW + (p/m)\partial_xW+m\gamma\partial_pW=0.
\label{eqq4}
\end{equation}
We look for the solution of this time-dependent equation using a scaling ansatz : $W(x,p,t)=W_0(X,P)$ with  $X = [x-\eta]/\alpha$ and $P = \alpha(p-m\dot{\eta})-m\dot{\alpha}(x-\eta)$ \cite{Castin}, 
where $\alpha$ and $\eta$ are time-dependent functions to be determined
with boundary conditions
\beq
\eta(0)=0,\;\;\dot\eta(0)=0,\;\; \alpha(0)=1,\;\;\dot\alpha(0)=0,
\eeq
so that $P(t=0)=p$ and $X(t=0)=x$. 
Let's calculate the derivative of $X$ and $P$ variables with respect to $x$, $p$ and $t$:
\begin{eqnarray}
&&\partial_t X= - (\dot \alpha X + \dot \eta)/\alpha \nonumber \\
&&\partial_x X= 1/\alpha\nonumber \\
&&\partial_t P = m\dot \alpha \dot \eta - m\alpha \ddot \eta + mX(\dot \alpha ^2 - \alpha \ddot \alpha) + \dot \alpha P/\alpha\nonumber \\
&&\partial_x P = -m\dot \alpha \nonumber \\
&&\partial_p P =\alpha.
\label{eqq6}
\end{eqnarray}
We deduce the following derivative of the phase-space distribution :
\begin{eqnarray}
&&\partial_t W= \partial_tX\partial_XW_0 + \partial_tP\partial_PW_0,
 \nonumber \\
&&\partial_x W= \partial_xX\partial_XW_0 + \partial_xP\partial_PW_0,
 \nonumber \\
&&\partial_p W=\partial_p P \partial_PW_0. 
\label{eqq7}
\end{eqnarray}
Combining Eqs.~(\ref{eqq7}) and (\ref{eqq4}), we get
\begin{equation}
\Lambda_X \partial_XW_0 + \Lambda_P \partial_PW_0=0
\label{eqfond}
\end{equation}
with
\begin{eqnarray}
\Lambda_X & = &(\partial_tX + (p/m)\partial_xX)= P/(m\alpha^2), \nonumber \\
\Lambda_P & = &(\partial_tP + (p/m)\partial_xP+m\gamma\partial_pP)\nonumber \\ & = &  -m\alpha (\ddot \eta-\gamma)
 - mX \alpha \ddot \alpha.
\label{eqq8}
\end{eqnarray}
The integration of Eq.~(\ref{eqfond}) over $X$ and $P$  yields the equation for the packet center $\ddot \eta=\gamma$. Multiplying Eq.~(\ref{eqfond}) by $XP$ and integrating over  $X$ and $P$ provides the equation fulfilled by the $\alpha$ scaling parameter:
\begin{equation}
\ddot \alpha=\frac{\langle P^2 \rangle_0}{m^2\langle X^2 \rangle_0} \frac{1}{\alpha^3} \equiv \frac{\omega_0^2}{\alpha^3},
\label{eqq14}
\end{equation}
whose solution is $\alpha(t)=(1+\omega_0^2t^2)^{1/2}$. 
For the ground state we find $\omega_0 = \hbar / (2m\sigma^2)$ where $\sigma=\Delta x_0$.

Let us now consider a Bose-Einstein condensate in the Thomas-Fermi regime. Initially the BEC is trapped in an harmonic trap of angular frequency $\omega_0$ and its wave function normalized to the number of atoms reads $\Psi(x,0)=n_0^{1/2}(x)$, where $n_0$ is the atomic density:
\begin{equation}
n_0(x)= (\mu-m\omega_0^2x^2/2)/ g,
\label{eqdens}
\end{equation}
where $g$ accounts for the strength of the interactions.
At time $t>0$, we propose to describe the dynamics using the ansatz
\begin{equation}
\Psi(x,t) = \frac{1}{\alpha^{1/2}}n_0^{1/2}\left( \frac{x-\eta}{\alpha} \right){\rm e}^{iS(x,t)}.
\label{eqansatz}
\end{equation}
To obtain the expression of the scaling parameter $\alpha$ and of the phase $S(x,t)$ we shall use the hydrodynamic equations. The continuity equation for the density $n(x,t)=|\Psi(x,t)|^2$ yields the expression for the velocity field \cite{shly}, 
\begin{equation}
\partial_tn+\partial_x(nv_x)=0 \Longrightarrow v_x=\dot\eta + \dot\alpha\frac{x-\eta}{\alpha}.
\end{equation}
The phase $S$ can be directly infered from the velocity field through the relation $v_x=\hbar\partial_xS/m$:
\begin{equation}
S(x,t)=\frac{m}{\hbar}\left( \dot\eta(x-\eta) + \dot\alpha\frac{(x-\eta)^2}{2\alpha}   \right).
\end{equation}
The Euler equation for the velocity field in the Thomas Fermi regime reads
\begin{equation}
m\partial_tv_x=\partial_x \left( -\frac{1}{2}mv^2 - V_{\rm ext}(x) -  g n(x,t)   \right).
\label{eqeuler}
\end{equation}
Combining Eqs.~(\ref{eqdens}), (\ref{eqansatz}) and (\ref{eqeuler}), we find the equations fulfilled by the center position of the packet $\eta$ and the one for scaling parameter $\alpha$, 
\begin{equation}
\ddot \eta=\gamma\qquad{\rm and} \qquad \ddot \alpha= \frac{\omega_0^2}{\alpha^2}.
\end{equation}
Instead of the $\alpha^3$ denominator that we found for the noninteracting gas in Eq.~(\ref{eqq14}),
we have now $\alpha^2$,  
so the solution cannot be written explicitly. However, as $\dot{\alpha}=\omega_0[2(1-\alpha^{-n+1})/(n-1)]^{1/2}$, the long time asymptotics, as $t\gg1/\omega_0$, is quite simple: 
$\dot\alpha\sim \omega_0$ for the noninteracting gas ($n=3$) and $\dot\alpha\sim \sqrt{2}\omega_0$ for the TF
regime ($n=2$). Once the mean field energy is released, the cloud expansion proceeds as 
a free expansion with a rate of change corresponding to an effective initial frequency $\sqrt{2}\omega_0$,
a fingerprint of the interactions that, remarkably, does not depend on $ g$.

The time-dependent Wigner function associated with $\Psi(x,t)$ can be recast in terms of the initial Wigner function:
\begin{eqnarray}
W(x,p,t) &  = &   \frac{1}{N\pi\hbar}\int \Psi^*(x+y,t)\Psi(x-y,t){\rm e}^{2ipy/\hbar}{\rm d}y 
\nonumber \\
 & =  &\frac{1}{N\pi\hbar\alpha}\!\int\!\!\left[ n_0\!\!\left(\! \frac{x+y-\eta}{\alpha}\! \right)\!n_0\!\!\left(\! \frac{x-y-\eta}{\alpha} \!\right)\! \right]^{\!1/2}
 \nonumber \\
&& \times {\rm e}^{2ipy/\hbar+iS(x-y,t)-iS(x+y,t)} {\rm d}y
\nonumber \\ & = & W_0\left(X,P\right). 
\end{eqnarray}
Interestingly enough, this result coincides with the one of the collisionless Vlasov equation \cite{dgo02}. This may appear surprising at first sight since Vlasov equation is a semiclassical equation in the presence of a mean-field term. However, the quadratic form of the density profile that enters the mean field potential term enables to conserve the equivalence between the equation of evolution for the Wigner function and the Vlasov equation. In this sense, this is a direct consequence of the initial harmonic confinement assumption.

The expression used above for the Wigner is the one associated with a pure state. This is valid in the absence of interactions and also for the mean field description of a Bose-Einstein condensate but it cannot be applied directly to a many-body wave function. In this latter case, the Wigner function is defined through the one-body reduced density matrix, $g_1(x,y;t)$ \cite{ADC08}:
\begin{equation}
W(x,p,t)   =    \frac{1}{\pi\hbar}\int g_1(x+y,x-y,t){\rm e}^{2ipy/\hbar}{\rm d}y. 
\end{equation}
A large class of many-body quantum systems exhibit self-similar dynamics \cite{Demler10,ADC11}. This includes the Calogero-Sutherland model \cite{Su98}, the Tonks Girardeau gas \cite{OS02, MG05}, certain Lieb-Liniger states \cite{BPG08}, Bose-Einstein condensate \cite{CD96, KSS96} even in the presence of dipolar interactions \cite{OGE04}, strongly interacting gas mixtures \cite{GM07}, ...
The self-similar dynamics can be written in our context as
\begin{equation}
g_1(x,y;t) = \frac{1}{\alpha}g_1(\frac{x-\eta}{\alpha},\frac{y-\eta}{\alpha};0)e^{i(S(x,t)-S(y,t))}
\end{equation}
yielding once again $W(x,p,t)   = W_0\left(X,P\right)$.

\section{$f$ functions}
We may calculate the $f_j$ functions in Eq.~(\ref{v1}) making use of Eqs. (\ref{ev1}-\ref{uxt}). 
An alternative route is to write down the Wigner function explicitly 
making use of the known phase-space propagator, 
\begin{eqnarray}
\!\!\!\!\!\!\!\!W(x,p;t) & = &  \int\!\!\int {\rm{d}}x_0 {\rm{d}}p_0 W(x_0,p_0;0)
\nonumber\\
&\times& \delta\!\left(\!x-x_0-v_0t-\frac{\gamma t^2}{2}\!\right)\! \delta(v-v_0-\gamma t)
\nonumber \\
&=&\frac{1}{\pi\hb} 
e^{-\frac{(p - m \gamma t)^2 2\sigma^2}{\hb^2}} e^{-\frac{(x - p t/m + \gamma t^2/2)^2}{2\sigma^2}}, 
\end{eqnarray}
for $W(x_0,p_0,0)=\frac{1}{\pi\hb} 
e^{-\frac{p_0^2 2\sigma^2}{\hb^2}} e^{-\frac{x_0^2}{2\sigma^2}}$. The end result is 
$W(x,p,t)=W(x_0,p_0,0)$ for phase-space points connected by classical trajectories (Liouville's theorem). 
Note that, in general, $x_0\ne X$ and $p_0\ne P$. An explicit calculation, however, shows 
the equality of this Wigner function with $W_0(X,P)$.  
\begin{widetext}
\beqa
%V_j(x, t):= 
 % {\rm e}^{-\Phi} f_j
f_1&=&
\frac{m \sigma t (8 \gamma m^2 \sigma^4 + \gamma \hb^2 t^2 + 2 \hb^2 x)}{
\sqrt{2\pi} (4m^2 \sigma^4 + \hb^2 t^2)^{3/2}},
\label{v1a}
\\
%V_2(x, t) &:=& 
f_2&=&
\frac{m \sigma \big\{64 \gamma^2 m^4 \sigma^8 t^2 + 
    16 \hb^2 m^2 \sigma^4 [\sigma^2 +  \gamma t^2 (\gamma t^2 + 2 x)] + 
    \hb^4 t^2 [4 \sigma^2 + (\gamma t^2 + 2 x)^2]\big\}}{
  2 \sqrt{2\pi} (4m^2 \sigma^4 + \hb^2 t^2)^{5/2}},
\label{v2a}
\\
%V_3(x, t) &:=& 
%{{\rm e}^{-\Phi}} f_3
f_3&=&
\frac{m \sigma t (8 \gamma m^2 \sigma^4 + \gamma \hb^2 t^2 + 2 \hb^2 x) }{4 \sqrt{2\pi} (4m^2 \sigma^4 + \hb^2 t^2)^{7/2}}
\nonumber\\
&\times&\Big\{ 64 \gamma^2 m^4 \sigma^8 t^2 + 
16 \hb^2 m^2 \sigma^4 [3 \sigma^2 + \gamma t^2 (\gamma t^2 + 2 x)] + 
\hb^4 t^2 [12 \sigma^2 + (\gamma t^2 + 2 x)^2]\Big\}.
\label{v3a}
\eeqa
\end{widetext}

\end{document}